\documentclass
[reprint,
superscriptaddress,
 amsmath,amssymb,
 aps,
 prl,
]{revtex4-1}
 \usepackage{color}
 \definecolor{MPPgreen}{RGB}{0,128,112}
\usepackage{graphicx}
\usepackage{dcolumn}
\usepackage{bm}

\newcommand{\addMPI}{\affiliation{Max-Planck-Institut f\"ur Physik,  D-80805 M\"unchen, Germany}}
\newcommand{\addTUM}{\affiliation{Physik-Department, Technische Universit\"at M\"unchen, D-85748 Garching, Germany}}
\newcommand{\addVNA}{\affiliation{Institut f\"ur Hochenergiephysik der \"Osterreichischen Akademie der Wissenschaften, A-1050 Wien, Austria, \\and Atominstitut, Vienna University of Technology, A-1020 Wien, Austria}}
\newcommand{\addPOR}{\affiliation{CIUC, Departamento de Fisica, Universidade de Coimbra, P3004 516 Coimbra, Portugal}}

\begin{document}

\bibliographystyle{unsrtnat} 

\title{Gram-scale cryogenic calorimeters for rare-event searches}	
\author{R. Strauss}\thanks{Corresponding author: strauss@mpp.mpg.de}
\addMPI
\author{ J. Rothe}
\addMPI
\author{G. Angloher}
\addMPI
\author{A. Bento}
\addPOR
\author{ A. G\"utlein}
\addVNA
\author{D. Hauff}
\addMPI
\author{ H. Kluck}
\addVNA
\author{M. Mancuso}
\addMPI
\author{ L. Oberauer}
\addTUM
\author{F. Petricca}
\addMPI
\author{F. Pr\"obst}
\addMPI
\author{J. Schieck}
\addVNA
\author{S. Sch\"onert}
\addTUM
\author{ W. Seidel}\thanks{Deceased 19 February 2017}
\addMPI
\author{ L. Stodolsky}
\addMPI





\date{\today}

\def\zz{neutrino }
\def\zzn{neutrino}
\def\zzs{neutrinos }
\def\zzsn{neutrinos}

\def\yy{energy }
\def\yyn{energy}
\def\yys{energies }
\def\yysn{energies}

\def\sctg{scattering }
\def\sctgn{scattering}

\def\csss{cross section }
\def\csssn{cross section}

\def\thld{threshold }
\def\thldn{threshold}

\def\cc{cryogenic calorimeter }
\def\ccn{cryogenic calorimeter}

\def\dr{detector }
\def\drn{detector}
\def\drs{detectors }
\def\drsn{detectors}

\def\temp{temperature }
\def\tempn{temperature}

\def\cy{crystal }
\def\cyn{crystal}

\def\bkgd{background }

\def\sc{superconducting }

\newcommand{\eq}[1]{Eq.\,\ref{#1}}

\begin{abstract}

The energy threshold of a cryogenic calorimeter can be lowered by reducing its size. 
This is of importance since the resulting increase in signal rate enables new approaches in rare-event searches, including the detection of MeV mass dark matter and coherent scattering of reactor or solar neutrinos. 
A scaling law for energy threshold vs. detector size is given. We analyze the possibility of lowering the threshold of a gram-scale cryogenic calorimeter to the few eV regime. A prototype 0.5\,g Al$_2$O$_3$ device achieved an energy threshold of $E_{th}=({19.7\pm0.9}$)\,eV,  the lowest value reported for a macroscopic calorimeter. 

\end{abstract}


\pacs{Valid PACS appear here}


\maketitle

\section{Introduction}
Cryogenic calorimetry \citep{physto} is based on the idea that the \temp rise in a target after an energy deposition $\Delta E$ is given by
\begin{equation} \label{jump}
\Delta T= \frac{\Delta E}{C}
\end{equation}
where $C$ is the heat capacity
of the object. A small $C$, which can be achieved in crystalline materials at $\sim$\,mK temperatures, 
leads to a large \temp jump and so to a high sensitivity to small \yysn. 
State-of-the-art cryogenic detectors  with a mass of $300$\,g
 have reached energy thresholds down to
$\sim300$\,eV~\citep{Angloher:2015ewa}.

\begin{figure}
\centering
\includegraphics[width=0.5\textwidth]{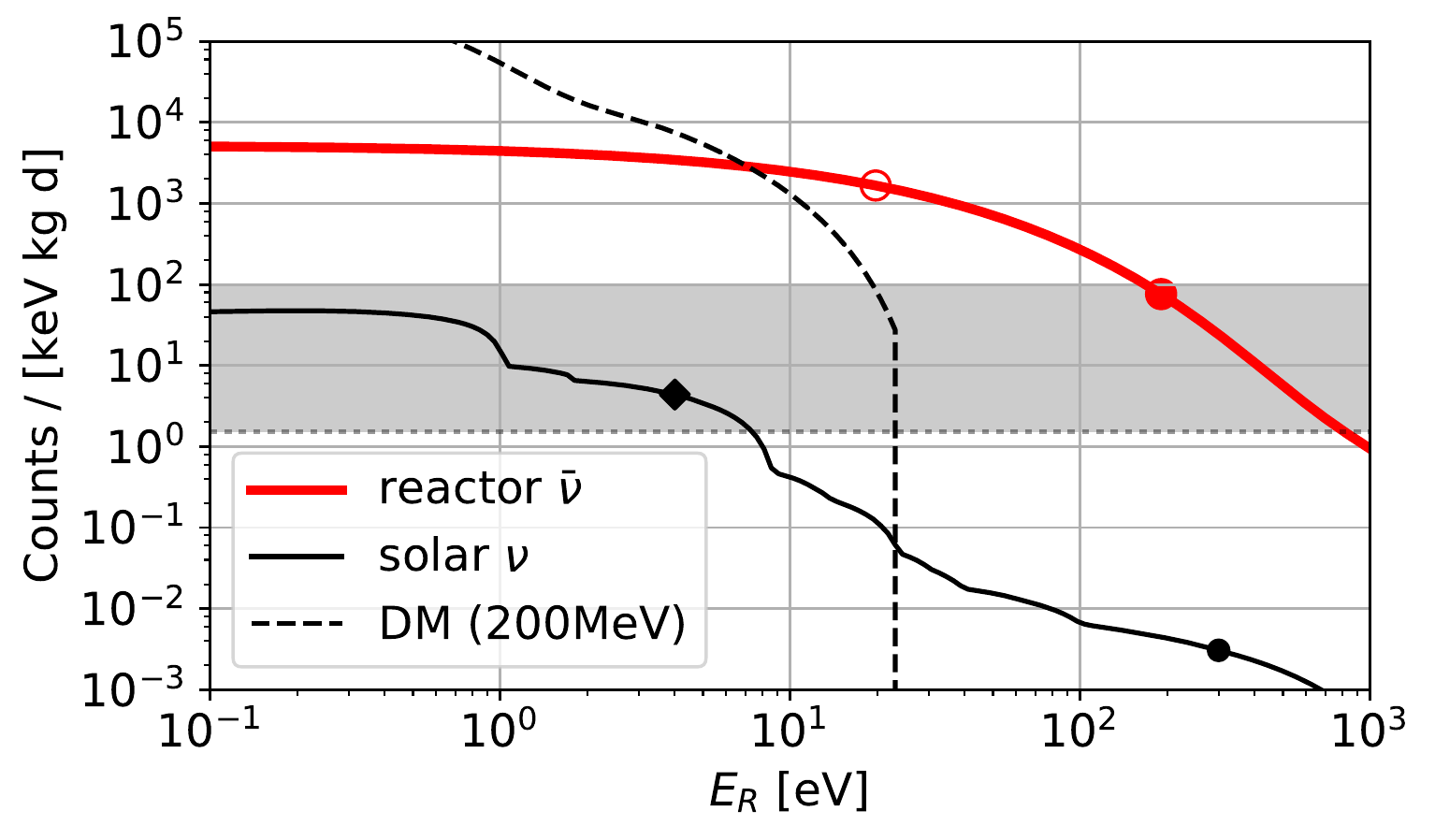}
\caption{Nuclear-recoil spectra on CaWO$_4$ 
for CNNS of anti-neutrinos from a 4\,GW nuclear reactor at a distance of 40\,m, 
for 200\,MeV/c$^2$ mass DM ($\sigma=1$\,pb), and 
for CNNS of solar neutrinos. The grey area indicates a range of measured background levels 
for shallow sites~\cite{Gastrich:2015owx,Akerib:2010pv} and at a deep-underground location (dotted line)~\cite{Strauss:2014aqw}. Closed circles indicate previous cryogenic detector technology~\cite{Angloher:2015ewa,Strauss2016}, the performance of the detector presented here is depicted as an open circle (surface operation). The expected  threshold of the presented detector at a deep-underground location is shown as  a diamond. 
}
\label{comp}
\end{figure}

 A further reduction of the \thld is of great interest since for
many important processes, such as coherent neutrino nucleus scattering (CNNS) 
or the \sctg of the hypothetical dark matter (DM) particles, the count
rate increases strongly as the \thld of a \dr is lowered. In Fig.\,\ref{comp}
we show the count rate in a CaWO$_4$ detector for various processes as a function of the recoil energy $E_R$: For CNNS of anti-neutrinos at a distance of 40\,m  from a 4\,GW nuclear power reactor (thick full line), for a hypothetical DM particle with a mass of 200\,MeV/c$^2$ and a cross-section of 1\,pb (dashed line), and for CNNS of solar neutrinos (full line).   
Measured background rates at different shallow low-background facilities~\cite{Gastrich:2015owx,Akerib:2010pv}, extrapolated to lower energies,  are shown as a grey band. The lower limit (dotted line) indicates the present fundamental background limitation due to the intrinsic radiopurity of CaWO$_4$ crystals, as measured deep underground~\cite{Strauss:2014aqw}.  One observes that a threshold in the 10\,eV regime can allow the count rates  to rise significantly above background levels. Below, we show that a rapid detection of CNNS at a nuclear reactor is in reach and that the technology offers unique possibilities for the detection of MeV-scale DM and solar neutrinos.

Since macroscopic amounts of target material are necessary to obtain 
reasonable count rates for such weak interaction processes, microscopic 
calorimeters with target masses below milligrams are impractical for rare-event searches.
Here we describe how it is possible to achieve a suitably low \thld
with a small but macroscopic calorimeter, and present
the results of a 0.5 g  prototype.

\section{Gram-scale cryogenic calorimeters}
 A straight forward application of \eq{jump} would seem to
suggest that an improved response for small \yys is to be achieved
by simply reducing $C$ by lowering the \temp or using smaller
devices~\citep{physto}.

 However, there are some subtleties in applying 
\eq{jump}. For example it may apply not to the whole device in
question, but to some sensitive subsystem. This explains why we
obtain results with macroscopic devices, on the order of grams in
mass, which one would have thought only possible with very small or
microscopic systems. Below we report an \yy \thld of $\sim20$\,eV
-- which is in the order of atomic binding energies -- for a 0.5 g device operated under
unshielded conditions.

\subsection{Pulse height}
To understand this apparently surprising result, 
we consider the operation of the \drs similar to those used in CRESST (Cryogenic Rare Event Search with Superconducting Thermometers) for direct dark matter search\,\citep{Angloher:2015ewa,Angloher:2012vn}.
These detectors operate out of thermal equilibrium after an energy deposition.
The signal originates from the initial 
ballistic non-thermal phonons of the particle
event occuring  in the absorber crystal. 
The deposited energy is measured using a superconducting transition-edge-sensor (TES).
This is explained at length in
Ref.\,\citep{Probst:1995fk}, and  we use the terminology and
notation of
this reference. In particular we deal with  \drs operating in the
calorimetric mode (section 3.3.1 in \citep{Probst:1995fk}), 
where the film thermalizes slowly with respect to the
signal duration and so integrates the energy of the 
incoming non-thermal phonons. 
This results in a temperature rise in the thermometer film given by
\begin{equation} \label{an}
\Delta T_\mathrm{film} =\epsilon \, \Delta E/C_{e}\,
\end{equation}
where $\epsilon$  is the fraction of the deposited \yy 
$\Delta E$ thermalized in the film, and $C_{e}$ is the heat
capacity of the electrons in  the film.

 Since  $\epsilon \, \Delta E $ is the \yy absorbed in the film,
\eq{an} amounts to  \eq{jump}, but applied
 to the electrons of the \sc film. Their \temp rise and thus our
readout signal is given by a pulse whose magnitude originates in
a microscopic system
, although the \dr itself is
macroscopic.

A fundamental limitation only arises through the fact that for 
very small $C$  a body coupled to a heat bath
has large irreducible \temp fluctuations 
 $(\Delta T/T)^2=1/C$\,\citep{moseley_xray}, where we use $k_{B}=1$. This corresponds to theoretical energy
resolutions of $\mathcal{O}$(1\,eV) for massive calorimeters with
masses of $\sim100$\,g \citep{formaggio}.
This is not a significant limitation for the \drs we consider (${C\sim10^8}$ for the W film used here), but
it does set a limit on an indefinite reduction of $C$.

\subsection{Scaling law}
A simple scaling law enables us to extrapolate results with previous CRESST \drs to smaller sizes. 
This is possible on the basis of \eq{an}. 
We are interested in the \thld \yy $E_{th}$, which is inversely proportional to the temperature rise $\Delta T_\mathrm{film}$ for a given energy deposition.

We arrive at the following approximate scaling law
for threshold energy vs. detector mass $M$:
\begin{equation} \label{eth}
E_{th}=(\mathrm{const.})\times M^{2/3}.
\end{equation}

The proportionality constant will depend on the material and geometry but not on the size of the crystal. Further, constant noise conditions and a similar performance of the TES sensors is assumed. 

\eq{eth} results from \eq{an}
as follows.   The appearance of   $\epsilon$,  representing the
fraction of the original non-thermal 
phonons thermalizing in the film shows that the main effect
entering in \eq{an} is the competition between the
thermalization in the \sc film and the thermalization on the
surfaces
of the crystal. We expect that  $\epsilon$  is small due to the
great difference in
the respective surface areas and is simply given  
 the ratio of thermalization rates which we write as $\kappa A$. Taking the ratio for the film and crystal surface one has
$\epsilon=
\frac{\kappa_\mathrm{film}A_\mathrm{film}}{\kappa_\mathrm{crystal}A_\mathrm{crystal}}$,
where the $\kappa$ refer
to material properties  and the $A$ to the surface areas.
Inserting in \eq{an} and keeping only factors related to
size, one obtains
\begin{equation} \label{tog}
\epsilon \frac{ \Delta E}
{C_{e}} = 
\frac{\kappa_\mathrm{film}A_\mathrm{film}}{\kappa_\mathrm{crystal}A_\mathrm{crystal}}
\frac{\Delta E}{C_{e}}\propto \frac{1}{A_\mathrm{crystal}}\,,
\end{equation}
where we have left out all factors that do not depend on the crystal size in the last step.
$A_\mathrm{film}$ has cancelled because  ${C_{e}\propto V_\mathrm{film} = A_\mathrm{film}\cdot h}$, and  the films all have about the same
thickness $h$. 
\eq{tog} shows that the main effect in reducing the size 
of the crystals is to reduce their surface area and so increase the
signal correspondingly.
For a cube one has $A_\mathrm{crystal}\propto d^2$ and $M\propto d^3$ where $d$
is the edge length. Since the threshold
varies inversely to the pulse height or $\Delta T_\mathrm{film}$, we obtain 
\eq{eth}.

\begin{figure}
\centering
\includegraphics[width=0.5\textwidth]{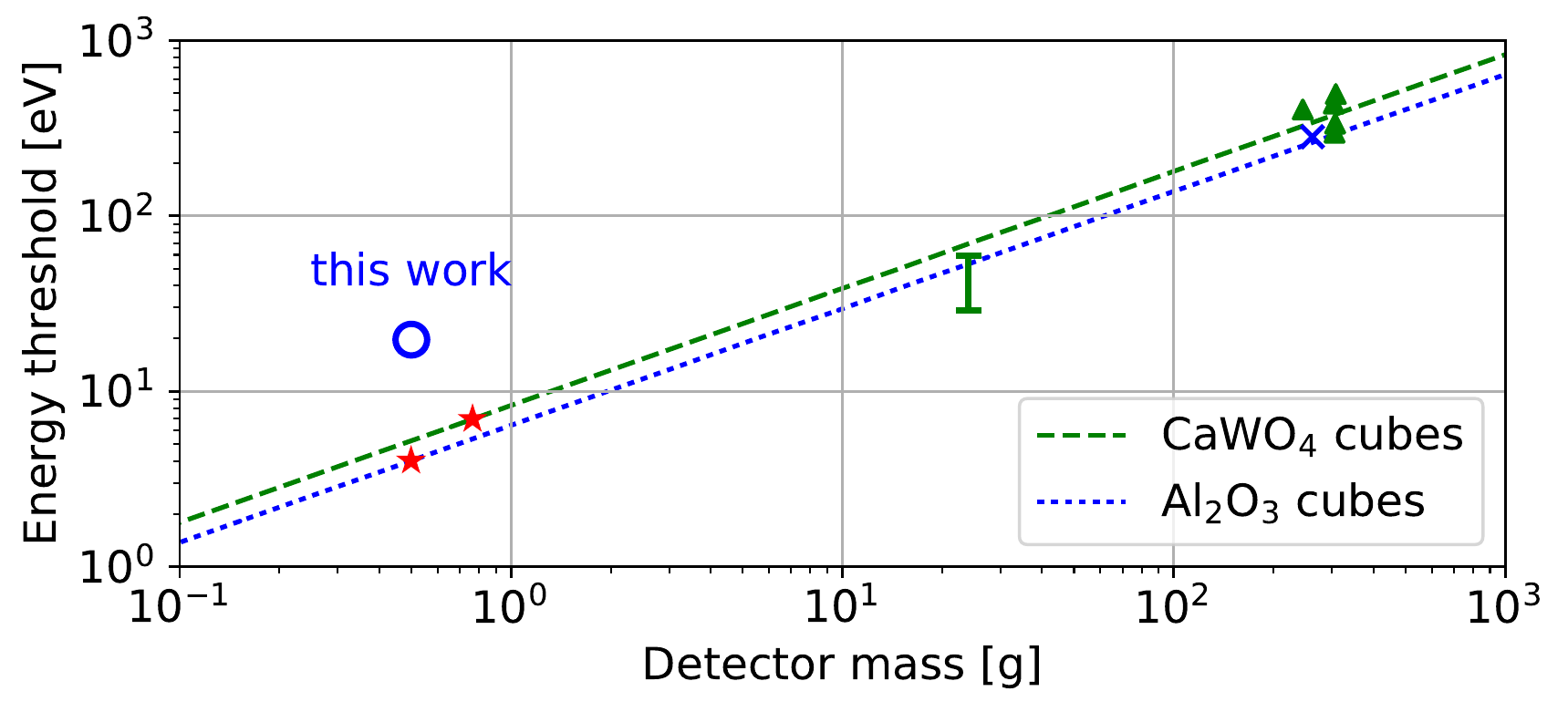}
\caption{ 
The scaling law for  CaWO$_4$  and Al$_2$O$_3$  cubes fitted to results of cryogenic  CaWO$_4$ (triangles) and Al$_2$O$_3$ (cross) detectors \citep{Angloher:2014myn,Angloher:2015ewa,Meier2000350}.   
The lines show the $M^{2/3}$ behavior according to \eq{eth}.  
The expected values for the prototype described here under low-background conditions is shown by the red stars. There is agreement with the expected performance of 24\,g CaWO$_4$ detectors (green error bar) \citep{Strauss2016}. 
The result presented here is shown by a blue circle.}
\label{fig:predictions}
\end{figure}

Fig.\,\ref{fig:predictions} shows the scaling law for CaWO$_4$ (dashed line) and Al$_2$O$_3$ (dotted line) calorimeters versus mass operated in the CRESST low-background setup, here referred to as benchmark setup. 
The offset  between the dashed and dotted lines is due to the different density and sound-speed of the materials and due to material-dependent transmission properties of phonons  into the \sc film (total factor $\sim$0.77). The model is fitted to the results of CaWO$_4$ detectors with $M=250-300$\,g (triangles) used in  CRESST-II  \citep{Angloher:2014myn,Angloher:2015ewa} and a
Al$_2$O$_3$ cube of $M=$ 262\,g  (cross) used in CRESST-I
\citep{Meier2000350}.  The scaling law nicely matches the  performance of  24\,g 
CaWO$_4$ detectors for CRESST-III  as infered from a above-ground measurement~\citep{Strauss2016}. The error bar indicates the  threshold range of this device when operated in the benchmark setup depending on the noise level of the individual readout channel~\cite{Strauss2016}.   
Red stars show the projected performance of ($5{\times} 5{\times} 5$)\,mm$^{3}$ CaWO$_4$ and Al$_2$O$_3$ calorimeters.  Below we discuss the result of a prototype 0.5\,g Al$_2$O$_3$ detector operated in a surface test setup (blue circle).

\section{Results from a prototype}

For the first test of a gram-scale calorimeter, an Al$_2$O$_3$ 
cube of ($5{\times} 5{\times} 5$)\,mm$^3$ with a mass of 
0.5\,g was used. All surfaces were optically
polished.  The cube was equipped with a W thin film TES 
 similar to those used for CRESST light detectors
\citep{Angloher2009270}. 
The TES design was adjusted for operation in the ``calorimetric mode''\,\citep{Probst:1995fk} on the cubic crystal.

The data presented here were acquired in a cryostat at the Max-Planck-Institut (MPI) for Physics in Munich, Germany.
The setup at surface-level had no shielding against environmental radio-activity or cosmogenic backgrounds.
The crystal was placed on three Al$_2$O$_3$ spheres with a diameter
of 1\,mm resting on a copper plate and providing point-like contacts. 
From the top, a bronze clamp (with a central Al$_2$O$_3$ sphere) presses on the cube. 
The electrical and thermal connections are realized by Al and Au wire bonds (diameter 25\,$\mu$m), respectively. 
The cryostat reached a base temperature of 6\,mK and the mixing chamber was stabilized during the measurement at 11\,mK.
The W  TES on the  Al$_2$O$_3$ crystal showed a normal-to-superconducting transition at 22\,mK. 
Commercial SQUID magnetometers and a state-of-the-art data acquisition system were used \citep{Angloher:2012vn}. 

A $^{55}$Fe X-ray calibration source with an activity of
0.6\,Bq was installed about 2\,cm from the Al$_2$O$_3$
cube. The detector was neither shielded against external radiation nor against radioactivity
originating from materials and surfaces inside the
experimental volume. 
The data amount to a total measuring time
of 5.3\,h  corresponding to an exposure of 0.11\,g-days. The total
particle pulse rate was 0.36\,Hz of which
$\gtrsim40$\,\% was from the X-ray lines of 
the $^{55}$Fe calibration source. 

The particle pulses can be well described by the thermal model for
cryogenic detectors \citep{Probst:1995fk}. A fit of the model to
the template pulse confirms that the Al$_2$O$_3$ detector is
operating in the calorimetric mode. The rise time, which
corresponds to the life time of the non-thermal phonons, is
$\tau_n=(0.30\pm0.01)$\,ms. The two decay times of the model,
 which depend on the thermal couplings of the TES and the absorber
crystal to the thermal bath, are found to  be 
$\tau_{fast}=(3.64\pm0.01)$\,ms and $\tau_{slow}=(28.17\pm0.09)$\,ms.

\subsection{Energy Calibration}

\begin{figure}
\centering
\includegraphics[width=0.5\textwidth]{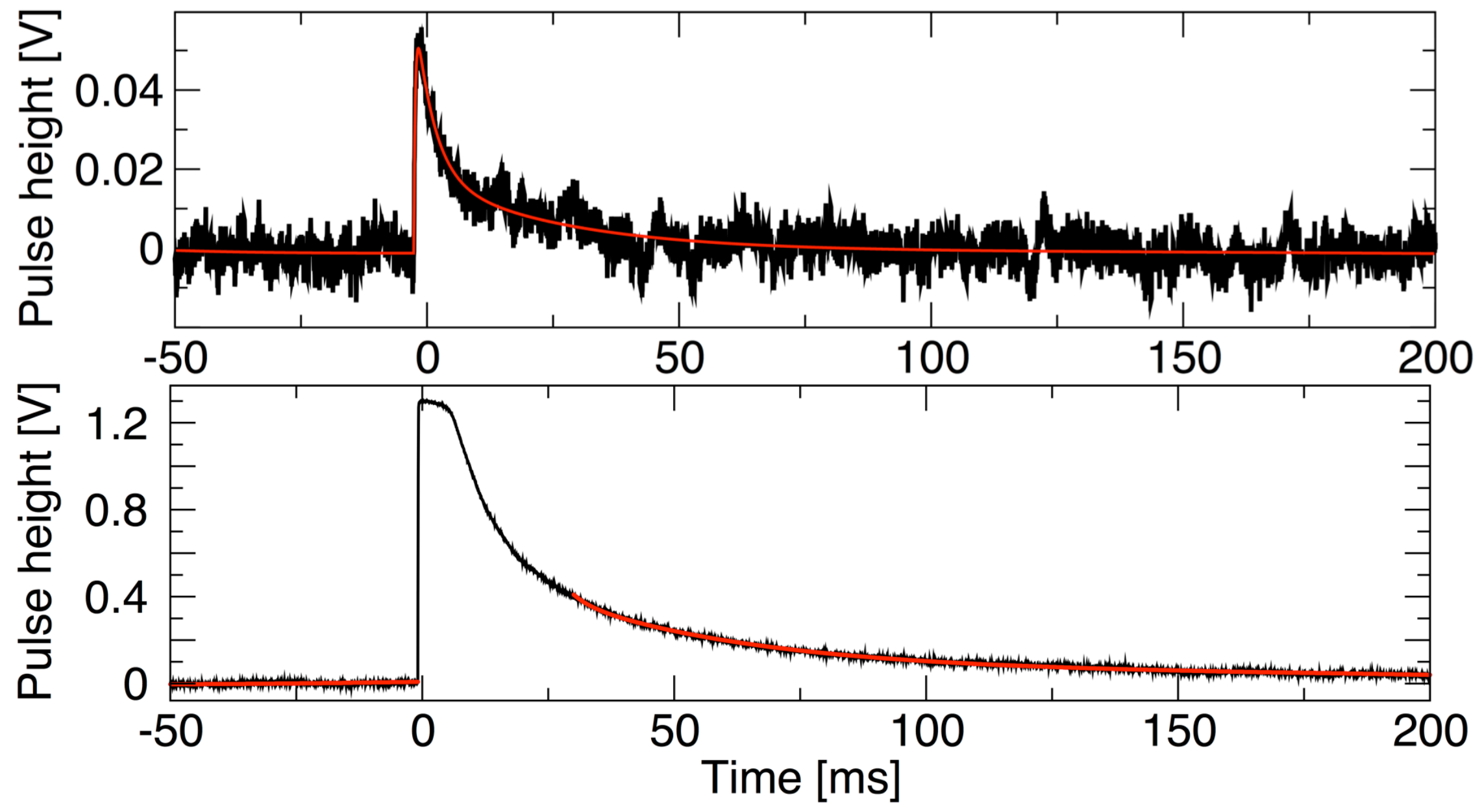}
\caption{ Randomly chosen pulses from different energy ranges fitted with the truncated template  (red lines). Upper frame: An event with an energy of about 100\,eV where the detector response is entirely linear. Lower frame: A saturated 5.9\,keV  pulse. Above 0.4\,V the pulse shapes deviate significantly from the template pulse.  }
\label{fig:fittedpulses}
\end{figure}

\begin{figure}
\centering
\includegraphics[width=0.5\textwidth]{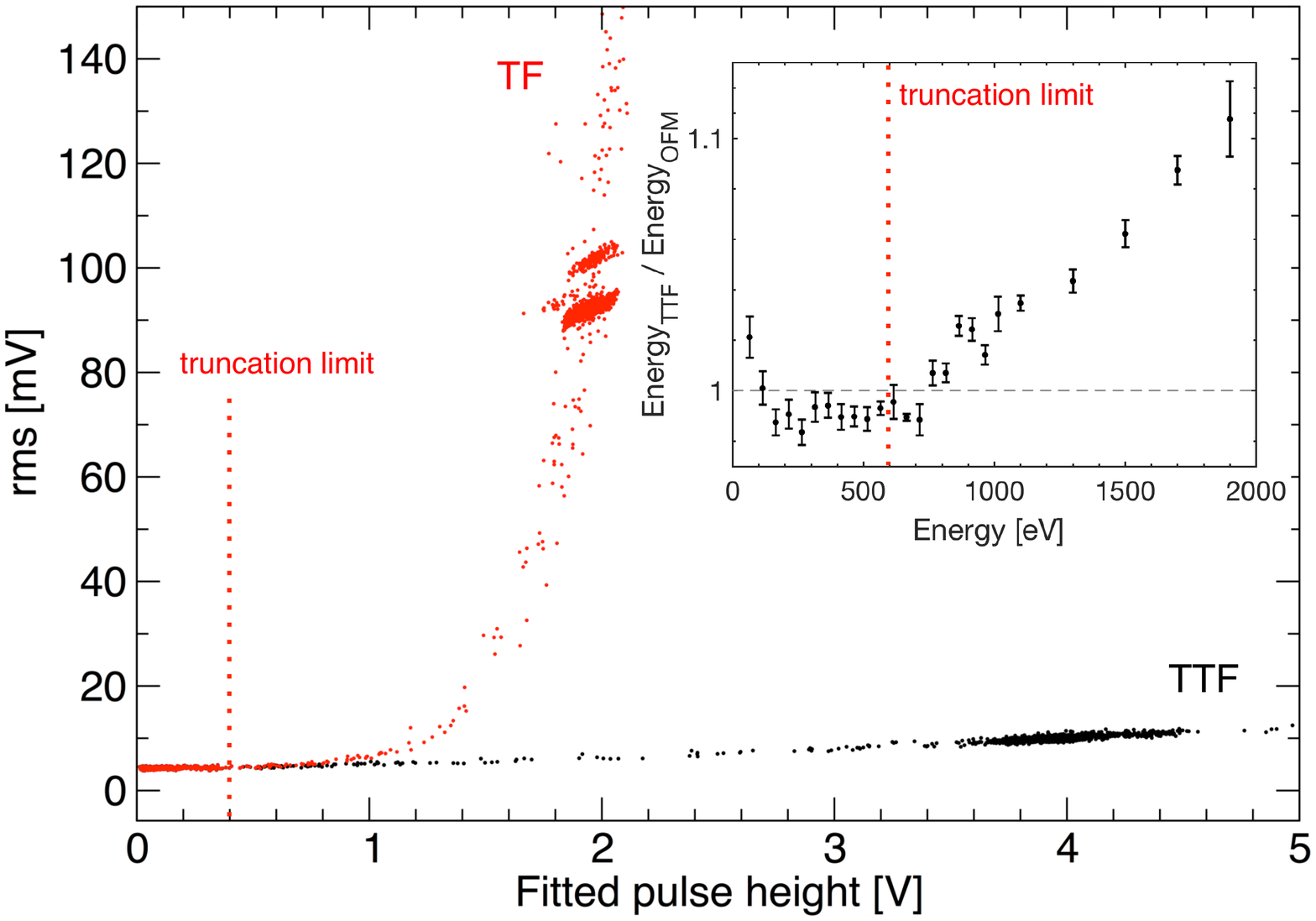}
\caption{Main frame: Goodness of the template fit (rms value). The truncation level is 0.4\,V and the truncated template fit (TTF) yields an almost energy-independent rms value (black). The $^{55}$Fe K$_\alpha$ X-ray line is reconstructed at a pulse height of (3.925$\pm$0.003)\,V.  The impact of pulse saturation on the energy reconstruction is demonstrated by a classic template fit (TF) without truncation (red). The rms value rises significantly for energies above the linear region. Inset: Ratio of the reconstructed energy by the TTF and the optimum filter method (OFM). In the linear region both methods agree on the 1\,\% level (see text). Above the truncation limit the OFM fails to reconstruct the correct energy, as expected, due to a different pulse shape caused by saturation.}
\label{fig:truncation}
\end{figure}

The energy of an event is inferred from its pulse height.
The pulses were fitted by a template pulse extracted from an energy
region were the pulse response is completely linear, in this case
around $\sim0.32$\,V, which corresponds to an energy of about
$0.47$\,keV.
The detector response becomes increasingly non-linear at about 3\,keV.
To reconstruct the energy of large, saturated, pulses the
method of a truncated template fit was used, a standard method in cryogenic calorimetry (see e.g. \citep{Angloher2009270}). The
pulse shape is fitted  in the linear region up to
the truncation limit, which here is chosen at 0.4\,V. Fig. \ref{fig:fittedpulses} (upper frame) shows a 100\,eV pulse  which is nicely fitted by the template pulse (red line). 
 In comparison, a 5.9\,keV pulse from the $^{55}$Fe source (lower frame) is  depicted. The event is only fit up to the truncation limit. Above this limit the pulse shapes deviate from that of the template. Fig. \ref{fig:truncation} (main frame) shows the fit goodness (rms value) of the truncated template fit (black) in comparison to a template fit without truncation (red).  The truncated fit shows no significant energy dependence of the rms values over the considered energy range. The slight rise (factor of $\sim2$) is expected since only part of the recorded pulse samples are exploited for the energy reconstruction. Without truncation, the fit fails above the linear region since the pulse shape of the events deviates significantly from the template. 
The dominant K$_\alpha$ line ($E_{lit}=5.895\,$keV) is
found to have a pulse height of (3.925$\pm$0.003)\,V and is
used for the calibration of the pulse spectrum.  The energy of the
$K_{\beta}$ line on the right shoulder is  then found at an energy of
$E_{K_\beta}=(6.485\pm0.017)$\,keV, which is in good agreement with
the literature value of 6.490\,keV. 

The energy reconstruction of the calibration line is robust against a change of the truncation limit. A variation of 20\% corresponds to a moderate error of 1.1\% on the energy calibration. This value is considered as a systematic error for the energy threshold (see below). Even a dramatic variation of the truncation limit by a factor of 2 changes the reconstructed pulse height of the X-ray lines by only $\sim3\,$\%.
 

\subsection{Energy spectrum}

\begin{figure}
\centering
\includegraphics[width=0.5\textwidth]{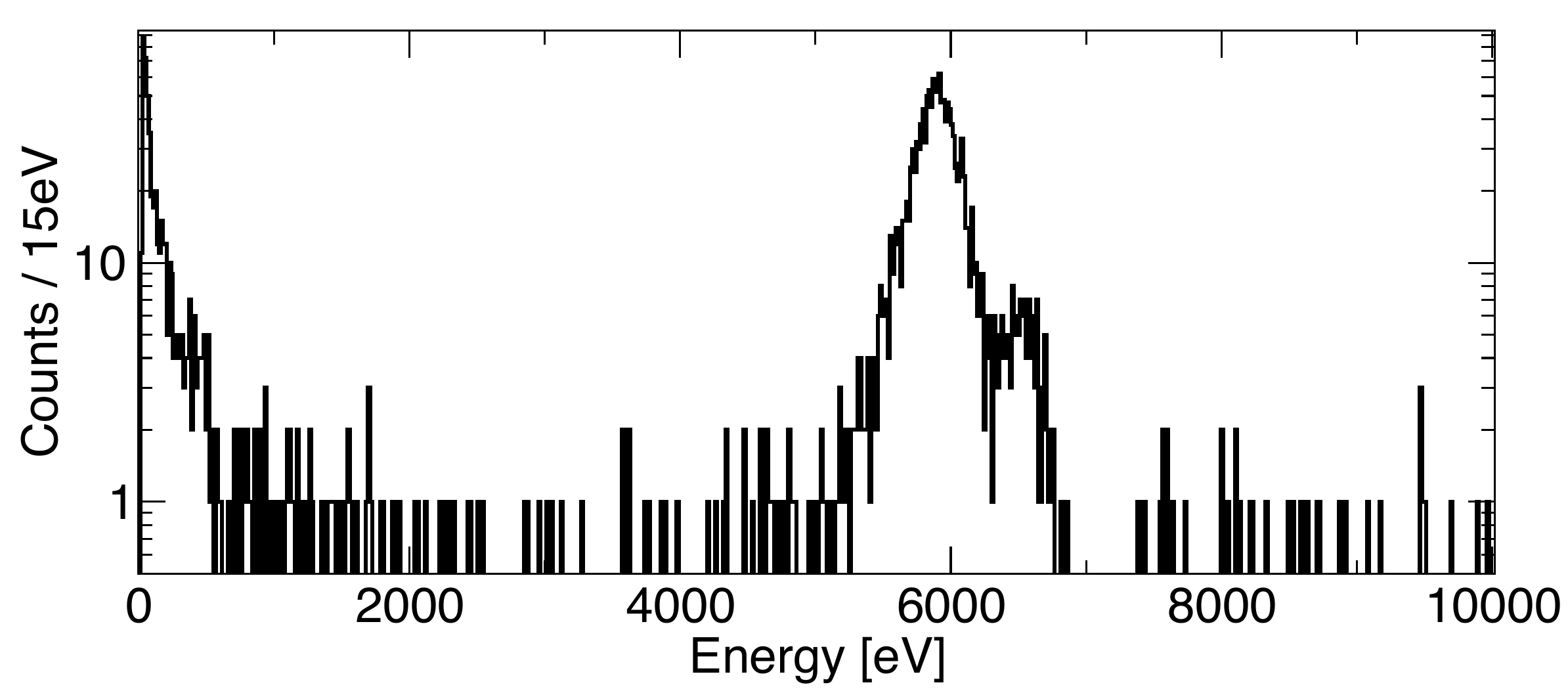}
\caption{ Final energy spectrum up to 10\,keV acquired with the
prototype Al$_2$O$_3$ detector in presence of a $^{55}$Fe
calibration source. }
\label{fig:spectra}
\end{figure}

Fig.\,\ref{fig:spectra} shows the final energy spectrum up to 10\,keV after stability and (standard) data quality cuts (see e.g.
\citep{Angloher:2015ewa,Angloher:2014myn}).
The energy resolution of the  reconstructed X-ray lines
is $\sigma_\mathrm{Fe}=(0.147\pm0.005)\,$keV which is significantly
larger than the baseline resolution at zero energy, as determined from 
the variance of the baseline (see below).  Part of this degradation (41\%) is due to the truncated fit that uses only partial information of  pulses at higher energies ($\gtrsim600$\,eV), as shown in Fig. \ref{fig:fittedpulses} (bottom).  
The remaining energy dependence of the detector resolution (by factor of 23 at 5.9\,keV) was observed 
previously for cryogenic Al$_2$O$_3$ detectors and matches the results of \,\citep{Sisti2001499}. 

The measurement shows a constant background rate  of 
$\sim1.2\cdot10^{5}$\,counts/[kg keV day] (7-10\,keV) above the
calibration peaks, see Fig.\,\ref{fig:spectra}. 
This background rate is not unexpected due to the lack of any shielding
against ambient and cosmogenic backgrounds.  
At lower energies ($\lesssim1$\,keV) the event rate significantly
rises towards threshold.
Auger electrons from the $^{55}$Fe source  
and surface-contamination induced backgrounds are the most
plausible explanation for this increase. 

Earlier results of cryogenic detectors operated in low-background setups, e.g. \citep{Angloher:2015ewa}, show a flat background on a level of 10\,counts/[kg~keV~day] (4-5 orders of magnitude lower compared to this measurement) down to the threshold energy of 300\,eV. This  clearly demonstrates that by a proper selection of the materials surrounding the detector, the surface background contribution can be drastically reduced. An active veto system by using Si slabs equipped with TESs which surround the calorimeter is planned. A dedicated Monte Carlo study shows that surface backgrounds can be reduced to negligible values~\cite{cnns_long}. 


\subsection{Threshold determination}

\begin{figure}
\centering
\includegraphics[width=0.5\textwidth]{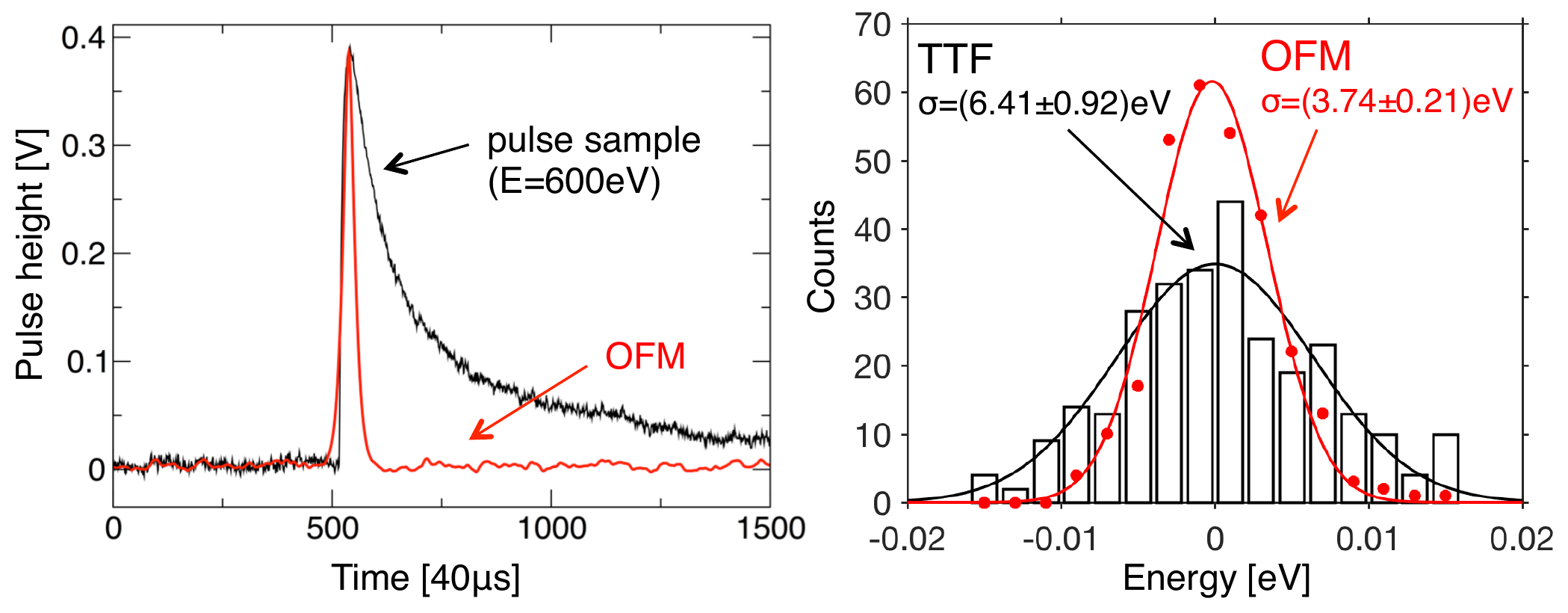}
\caption{Left: Pulse sample and optimum filter output in the time domain. Right: Comparison of the baseline noise derived by the truncated template fit (TTF, black histogram) and the optimum filter method (OFM, red dots). Gaussian fits to the data are shown. }
\label{fig:baselineOpt}
\end{figure}

For the pulse height evaluation at low energies, the optimum filter method is used 
(see e.g. \citep{Gatti:1986cw,Piperno:2011fp}). 
The optimum filter  maximizes
the signal-to-noise for a known signal, in our case the
template pulse, in the presence of stochastic noise with a known 
power spectrum. To build the filter transfer function, the Fourier
transform of the template pulse and the noise power spectrum are
required. Here, the latter is derived from $\sim400$ randomly
chosen baseline samples. In the selection of the baseline samples
the same quality cuts were applied as for the pulse samples. In the
frequency domain, the optimum filter
weights the spectral components according to their
signal-to-noise
ratio. Usually the filter is applied in frequency space to
minimize the computing time and is then transformed back to the 
time domain.  The result is normalized so
that it reproduces the unfiltered pulse height at 
the pulse's maximum (see Fig. \ref{fig:baselineOpt}, left). The energy reconstruction by the optimum filter agrees with that of the truncated template fit on a 1\%-level in the linear region (up to 600\,eV) and deviates significantly above as expected due to a different pulse-shape caused by saturation (see Fig. \ref{fig:truncation}, inset).   Below the truncation limit, a maximal deviation of 2.8\,\% is observed which is considered as systematic error of the energy calibration.
The baseline energy resolution after filtering is found 
to be 
$\sigma_b=(3.74\pm0.21)$\,eV.  
This compares to a value of $(6.42\pm0.92)$\,eV without filtering, showing a clear improvement (see Fig. \ref{fig:baselineOpt}, right).
Accordingly, this effect  reduces the energy threshold. This improvement can be exploited  using a data acquisition system which continuously streams the detector output, so that the pulse-triggering can be done in post-processing, when signal and noise power spectra are known.

\begin{figure}
\centering
\includegraphics[width=0.5\textwidth]{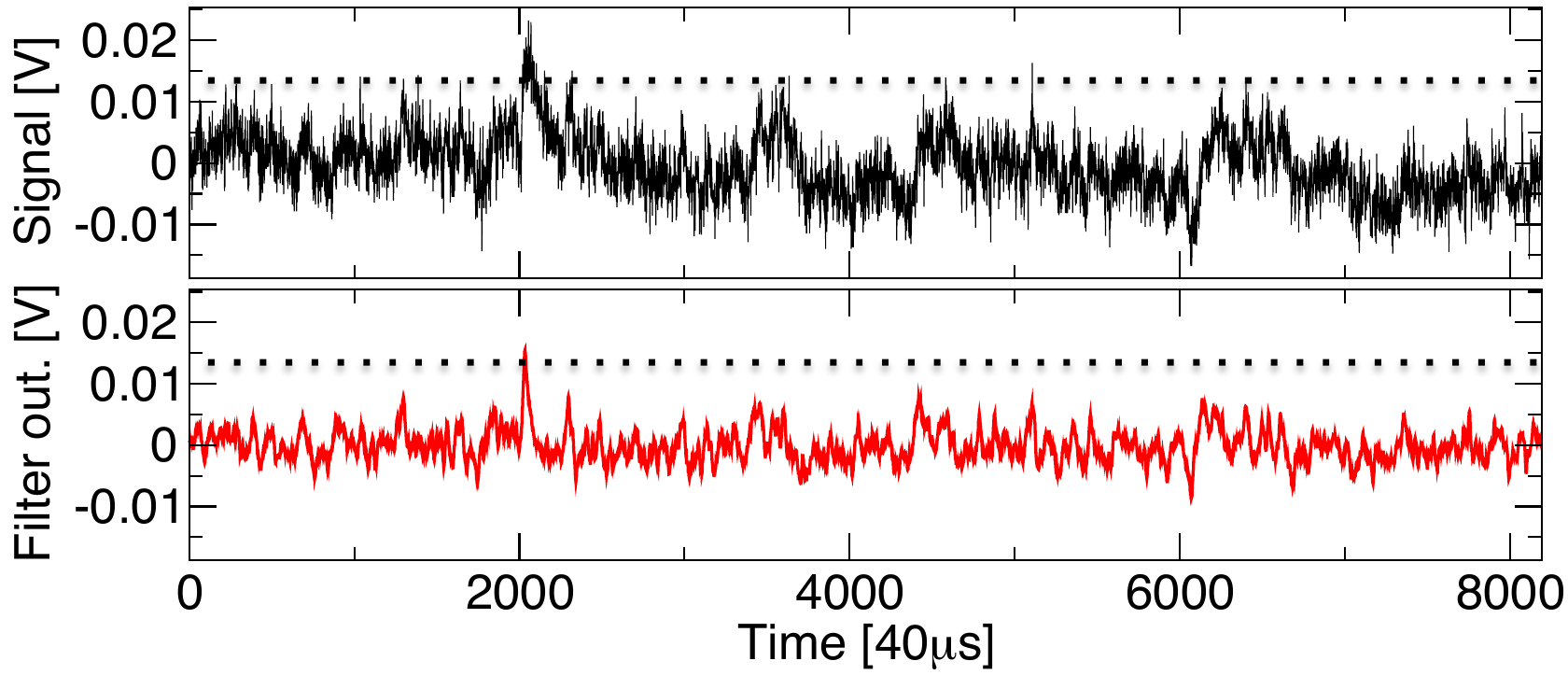}
\caption{Demonstration of the optimum trigger. Upper plot: A
19.7\,eV standard pulse is superimposed on a randomly chosen noise
sample (onset at sample 2000). Lower plot: Output of the optimum
filter applied to the sample. The pulse is clearly triggered while
noise contributions are  suppressed sufficiently below threshold which is set at a pulse
height of 13.0\,mV (see~text). }
\label{fig:trigger_demo}
\end{figure}

The functionality of such a software trigger based on the optimum filter is illustrated in Fig.\,\ref{fig:trigger_demo}. A small artificial pulse is superimposed on
a randomly selected baseline sample, drawn in the upper frame. The
lower frame shows the optimum filter output. The artificially added pulse clearly is 
seen above a given threshold (dotted line), while the random noise, which has a different 
pulse shape, is suppressed. 

\begin{figure}
\centering
\includegraphics[width=0.5\textwidth]{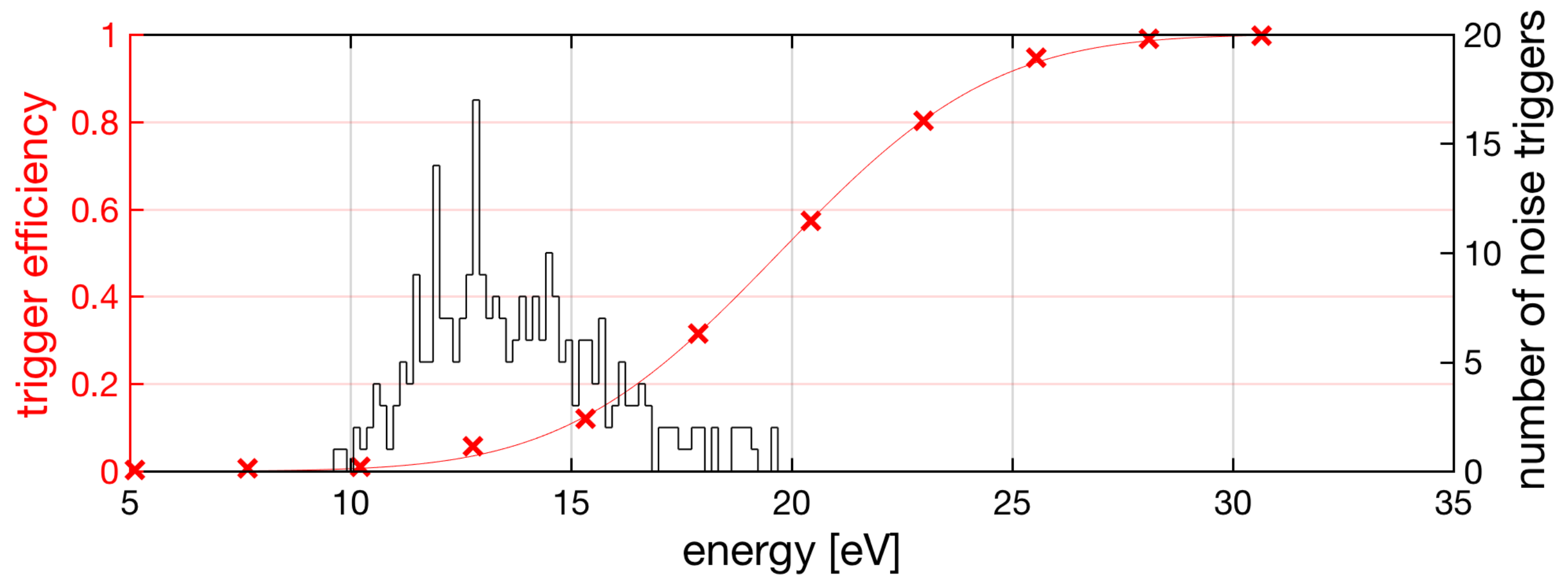}
\caption{Determination of the trigger threshold. Randomly chosen
noise samples are superimposed with template pulses of different
discrete energies (red crosses).  The optimum trigger is applied to
these samples yielding the energy-dependent trigger efficiency
(left y-axis). The data is fitted by an error-function, giving
an  energy threshold of  $E_{th}=(19.7\pm0.1)$\,eV
for 50\% efficiency. The width  $\sigma_{th}=(3.82\pm0.15)$\,eV is 
in agreement with the 
variance of the baseline noise. 
The reconstructed energy of pure noise samples after filtering is shown in a histogram (black, right y-axis).}
\label{fig:threshold}
\end{figure}

In the following we discuss how the energy threshold and the trigger efficiency can be determined in a direct way. 
Generally speaking, the threshold on the output of the optimum filter has to be chosen so 
as to be sensitive to the smallest possible energy depositions, while 
at the same time suppressing noise triggers sufficiently.
Fig.\,\ref{fig:threshold} (histogram, right axis) shows the
filter output of a set of pure noise samples. In contrast to the determination of the baseline noise (see above), the pulse position (in time) is not fixed but the algorithm runs over the noise trace and returns the maximal filter output. This explains the positive average reconstructed energy. 
The bulk of the noise samples have a reconstructed energy between 10 and 15\,eV with a tail up to $\sim19\,$eV. Most probably the latter is due to small pulses on the noise samples which can not by identified by data-quality
cuts selecting the noise samples. This effect is enhanced due to the exponentially increasing rate towards threshold in this calibration measurement.
It is reasonable to set the trigger threshold just above this assumed noise population.

We choose a trigger threshold of 13.0\,mV and validate this choice by a study 
of the trigger efficiency as a function of energy.
Onto the set of baseline samples, template pulses of various discrete
pulse heights (from about 1 to $10\cdot\sigma_b$) are added. 
The energy dependent trigger efficiency is  the fraction of the filtered artificial pulse samples  which fall above the threshold. 
Fig.\,\ref{fig:threshold} (left axis) shows the results of this
procedure for the discrete pulse heights (crosses). The resulting curve
can be nicely fitted by the function
$p_{trig}(E)=0.5\cdot(1+\mathrm{erf}[(E-E_{th})/(\sqrt{2}\sigma_{th})]$, 
where erf is the Gaussian error function
\citep{Angloher:2014myn}. 
The validity of the threshold choice manifests itself as a vanishing trigger efficiency at low energies,
corresponding to negligible noise triggers.
Furthermore, the width of the error function is 
$\sigma_{th}=(3.83\pm0.15)$\,eV which is in good agreement with
the baseline noise of $\sigma_b=(3.74\pm0.21)$\,eV, demonstrating
that the resolution of the detector at these energies is dominated
by the baseline noise. 
The energy threshold (by definition at 50\% trigger efficiency) is
found to be $E_{th}=(19.7\pm0.1(\mathrm{stat.}))$\,eV, which corresponds to $5.27\sigma_b$.
In a setup with optimized noise conditions, the threshold can presumably be lowered to
4.5-5$\sigma_b$, a typical value reached in low-background underground environments~\citep{Angloher:2014myn}.  Considering the systematic errors of the energy calibrations (see above), the energy threshold is $E_{th}=(19.7\pm0.9)$\,eV.

\subsection{Discussion of results}

The prototype detector shows the lowest energy
threshold reported for macroscopic calorimeters, an improvement by
one order of magnitude with respect to previous results\,\citep{Strauss2016}. 
It should be stressed that, since the device is a calorimeter, its calibration does not rely on quenching factors 
that arise when dealing with ionization or scintillation \drsn. The detector 
has the same response for a given \yy deposit, regardless
 of the interacting particle type and can approach energies below the fundamental nuclear-recoil reach of ionization detectors (e.g. 40~eV for
Si and Ge\,\cite{Agnese:2016cpb}).
The threshold, shown as a blue circle in Fig.\,\ref{fig:predictions},  is  higher than predicted by the scaling law, indicated by the red stars. 
However, this is expected to improve in a low-noise, low background environment.
When operated in a low-noise underground setup, the energy threshold of the investigated Al$_2$O$_3$ detector is expected to improve by a factor of 1.5-3\,\citep{Strauss2016}, to a value between 6 and 13\,eV. Further improvements (e.g. optimizing the TES design) are foreseen to  fully match the expectation of $4\pm1$\,eV.  


The threshold is determined by comparing the measured noise with
the amplitude of template pulses calibrated with events induced by
the $^{55}$Fe source. Thus  the main assumption in arriving at our
very low threshold is in the linear extrapolation of the pulse
height-energy relation to low \yysn. Our 
 threshold lies in a new energy range, never explored with this
type of detector. Nevertheless, such low energies still correspond  to the
creation of very many of the  $\mathcal{O}$(1\,meV)  non-thermal phonons involved
in the detection mechanism \citep{Probst:1995fk}.  For  energy
depositions around or below 10\,eV there is the possibility of
different nuclear recoil mechanisms as one approaches lattice
dislocation energies.
It could be interesting for rare event searches to attempt a direct
calibration at lowest energies, as with recoils from neutron capture, nuclear
isomeric transitions or electron capture.

\section{Outlook for rare-event searches}

Due to the smallness of the calorimeters presented here, the
technology permits new experiments in three aspects:
1) ultra-low energy thresholds down to the 10\,eV-regime, 
2) encapsulation of the small calorimeters by cryogenic veto
detectors and 
3) ability to operate the detectors above-ground in a high-rate
environment.
These features enable interesting possibilities involving the
exploitation of the enhancement of the cross section by coherent
scattering\,\citep{Drukier:1983gj}.

A new energy regime of nuclear recoils is accessible for the first time with this technology which implies that a new range of DM particles can be probed. The prototype run
described here can be used to set a new limit on the
spin-independent DM particle-nucleon cross section $\sigma_{\textrm{SI}}$ for
masses below $m_{\textrm{DM}}=500$\,MeV/c$^2$\,\citep{MeV_long}, extending the reach of direct DM search experiments down to $m_{\textrm{DM}}=150$\,MeV/c$^2$. Operated in a low-background setup gram-scale detectors will significantly improve in sensitivity. Assuming the present detector performance using Al$_2$O$_3$ and a background level of 10\,counts/[kg keV day], which seems feasible even at a shallow site~\cite{Gastrich:2015owx},  an upper limit for  $\sigma_{\textrm{SI}}$ of $\sim10^{-5}$\,pb (at 500\,MeV/c$^2$) can be achieved with a moderate exposure of 1\,kg-day. Improvements on the threshold, as predicted by the presented scaling law, might enable searches for DM masses in the 10\,MeV regime.

The rate of solar neutrinos scattering coherently on nulcei in CaWO$_4$ exceeds the present intrinsic background level at recoil energies below $\sim7$\,eV, as shown in Fig. \ref{comp}. In this energy regime, the spectrum is dominated by scatters from Be$^7$ and pp neutrinos of the solar cycle. An array of gram-scale detectors with a total target mass of $\mathrm{O}$(1\,kg) would allow flavor-independent precision measurements of the solar neutrino flux and enable new solar physics \cite{Cerdeno:2016sfi}.


The  technology presented here shows promise for a rapid detection of coherent neutrino nucleus scattering (CNNS) and opens a window to study many interesting physics scenarios beyond the Standard Model \citep{Lindner:2016wff}.  Lowering the energy threshold from, e.g. 300\,eV  to 20\,eV,  boosts the expected count rate by about two orders of magnitude (Fig. \ref{comp}).   
Operating an array of such detectors with a total mass of 10\,g  at a distance of 40\,m from a GW-scale nuclear power reactor, yields a count rate of $\sim10^3$\,counts/[kg keV day] which is a factor of 10-10$^3$ above expected backgrounds (see Fig. \ref{comp} and \citep{cnns_long}).  A 5$\sigma$ discovery of CNNS is then expected within $\lesssim2$\,weeks of measuring time, and precision measurements of the cross-section are in reach. Such a small-scale experiment can be realized with a commercial cryostat, standard lab electronics and a (presumably) compact shielding, which results in a moderate cost estimate. 
With this technology, real-time monitoring of nuclear reactors for non-proliferation and accident control is in reach.


\bibliography{CNNS_letter}

\end{document}